\documentstyle[12pt]{article}

\def\lapprox{\mathrel{\mathop {\hbox{\lower0.5ex\hbox{$\sim$}\kern-
0.8em\lower-0.7ex\hbox{$<$}}}}}
\def\gapprox{\mathrel{\mathop {\hbox{\lower0.5ex\hbox{$\sim$}\kern-
0.8em\lower-0.7ex\hbox{$>$}}}}}

\def\GN{G_{\rm N}}
\def\anti{\overline}
\def\eV{{\rm eV}}

\def\mn#1(#2){#1{\times}10^{#2}}
\def\YP{Y_{\rm P}}

\begin{document}
\hbox to\textwidth{November 1995\hfil MPI-PTh/95-115}
\bigskip\bigskip
\begin{center}
{\Large The Nature of Dark Matter\footnote{To be published in {\sl The
Evolution of the Universe}, edited by G.~B\"orner and S.~Gottl\"ober, Dahlem
Workshop Report ES~19 (John Wiley \&\ Sons).}}

\bigskip
\bigskip

Georg Raffelt\\
Max-Planck-Institut f\"ur Physik\\
F\"ohringer Ring 6, 80805 M\"unchen, Germany

\end{center}

\bigskip

\begin{abstract}
\noindent The dynamics of the universe may be dominated by novel weakly
interacting elementary particles, by baryons in an invisible form, by black
holes, and globally by vacuum energy. The main arguments for and against such
hypotheses are reviewed.
\end{abstract}

\bigskip


\section{Introduction}

On galactic scales and above, the mass density associated with luminous matter
(stars, hydrogen clouds, x-ray gas in clusters, etc.) cannot account for the
observed dynamics on those scales (Trimble 1987), revealing the existence of
large amounts of dark matter (DM) or else pointing to a breakdown of Newtonian
dynamics or the conventional law of gravity. The role of DM could be played by
anything from novel weakly interacting elementary particles to normal matter
in some invisible form (small stars, black holes, molecular gas, etc.). A
global, homogeneously distributed DM component can be provided by the vacuum
in the form of a cosmological constant. It is attempted here to summarize the
main arguments that have been advanced for and against various solutions to
the DM problem.

\newpage

\section{Big-Bang Nucleosynthesis}

One of the standard arguments for the existence of nonbaryonic DM is provided
by big-bang nucleosynthesis (BBN). In the standard big-bang scenario of the
early universe, the primeval abundance of the light isotopes $^2$H, $^3$He,
$^4$He, and $^7$Li has been produced after the freeze-out of $\beta$
equilibrium between protons and neutrons (B\"orner 1992; Peebles 1993). The
large number of photons relative to baryons keeps the nuclei in a dissociated
phase for a relatively long time after $\beta$ freeze-out, allowing neutrons
to decay freely during that epoch. Therefore, the yield of these isotopes is
very sensitive to the number ratio $\eta$ between baryons and photons. It is
noteworthy that $\eta$ remains the only free parameter of standard BBN after
the neutron lifetime and the total number of neutrino flavors ($N_\nu=3$) have
been measured in laboratory experiments with high precision (Particle Data
Group 1994). Because the present-day number density of cosmic microwave
background (CMB) photons is well measured one can translate $\eta$ into a
value for the cosmic baryon density in units of the cosmic critical density,
$\Omega_B h^2=0.0037\,\eta_{10}$ where $\eta_{10}=\eta/10^{-10}$. Here, $h$ is
the present-day Hubble expansion parameter in units of
$100\,\rm km\,s^{-1}\,Mpc^{-1}$.

BBN must confront two all-important questions: Are the observationally
inferred primeval light-element abundances consistent with a single $\eta$
value? If so, does it imply that some of the dynamical mass of the universe
must be nonbaryonic? The answer to the first question is the subject of a
renewed debate that has not come to a final conclusion at the time of this
writing (e.g.\ Copi, Schramm, and Turner 1995; Hata et~al.\ 1995).

The most important observational indicators for $\eta$ are the primeval
deuterium (D) and $^4$He abundance while $^7$Li does not play a particularly
decisive role because the BBN predictions as well as the observational data
are bedeviled by large uncertainties. The primeval $^4$He abundance inferred
from metal-poor HII regions is $\YP=0.232\pm0.003$ (statistical error) with a
systematic uncertainty estimated by different authors between $\pm0.005$
and $\pm0.014$. Based on the former, Hata et~al.\ (1995) infer a 95\% CL range
$\eta_{10}=1{-}3$.

The primeval D abundance used to be derived from the local interstellar
medium, and from abundance measurements of $^3$He together with a chemical
evolution model for D and $^3$He, the main idea being that all of the D
destroyed in stars should show up in the form of $^3$He. For example, Hata
et~al.\ (1995) infer a primeval D/H number ratio of $1.5{-}5.5{\times}10^{-5}$
which leads to a 95\% CL range $\eta_{10}=4.5{-}9$ which does not overlap with
the helium-inferred $\eta$ range. On the other hand, recent measurements of D
absorption lines in the spectrum of a high-redshift quasar yields a D
abundance of around $10^{-4}$ which correspond to $\eta$ values consistent
with those inferred from $^4$He (Songaila et al.\ 1994; Carswell et al.\ 1994;
Rugers and Hogan 1995). While the D absorption lines, in principle, could be
an artifact caused by a hydrogen ``interloper cloud'' which is located at a
suitable redshift, this interpretation may not fit the characteristics of the
observed line shape. Confusingly, from D absorption lines in a different
quasar system Tytler and Fan (1994) find a low D/H ratio which conforms to the
traditional interstellar medium ones.

If the high absorption-line values for D are correct, the $^3$He-inferred
values must be plagued by unaccounted systematic effects, probably from overly
simplistic assumptions concerning the chemical evolution of $^3$He. The $\eta$
value would be consistent with that inferred from $^4$He, indicating a low
baryon density of $\Omega_B h^2=0.004{-}0.01$ which would be just right to
account for all of the luminous matter, plus some dark baryons to spare, but
far from enough to provide all the required dynamical mass of $\Omega_{\rm
dyn}\gapprox0.2$. Therefore, most of the cosmic mass density would have to
consist of some novel form of matter, probably some weakly interacting
elementary particles.

On the other hand, if the low D values are correct, the minimal BBN scenario
is only consistent if one appeals to large systematic uncertainties, perhaps
in the interpretation of the observational data, perhaps in the details of
BBN. There are many ways in which the minimal or standard picture of BBN could
be incomplete; anything from decaying neutrinos to inhomogeneous baryon
distributions could modify the standard predictions (Malaney and
Mathews~1993). While the more conservative extensions of minimal BBN do not
seem to allow easily for a large enough baryon density to account for the
known mass in the universe, it is impossible to exclude this option if one is
willing to espouse a sufficiently radical nonstandard BBN scenario.

In summary, within standard BBN together with current observational data a
dominant nonbaryonic DM component is strongly suggested. Still, the correct
value of the primordial deuterium abundance remains unsettled so that the
consistency of standard BBN is not entirely assured. Therefore, it may be
premature to rule out a purely baryonic universe on the basis of BBN alone,
especially if the overall matter density is as low as 0.2 or 0.3 of the
critical density, a possibility which is suggested by the current best-fit
parameters of a Robertson-Walker-Friedmann-Lema\^\i tre model of the~universe.

\section[...]{Candidate Particles and Search for Their Existence}

\subsection{Neutrinos}

If most of the cosmic DM does not consist of baryons, what else can it be?
Obvious possibilities are primordial black holes or weakly interacting
elementary particles. The latter would need to carry a nonvanishing mass,
would need to be stable over cosmological times, and must have been left over
from the early universe with the right relic abundance. At the present time,
no particle with the requisite properties is known. The most popular
conjectures include neutrinos with mass, the lightest supersymmetric particle
(probably the neutralino), and axions.

Beginning with massive neutrinos, it is noteworthy that all three neutrino
flavors $\nu_e$, $\nu_\mu$, and $\nu_\tau$ must obey the cosmological mass
limit $m_\nu\lapprox 30\,\eV$ unless one postulates a fast decay channel
beyond what is predicted by the particle-physics standard model. The cosmic
neutrino sea should encompass around $100\,\rm cm^{-3}$ neutrinos of each
flavor. If one of them had a mass in the $10\,\eV$ range it would
significantly contribute to~DM.

No method is known that would allow one to measure the cosmic neutrino sea,
whether or not neutrinos have a mass and whether or not they cluster on
galactic scales. The only realistic chance to detect neutrino masses in the
cosmologically relevant range is to observe neutrino oscillations. The solar
neutrino problem is currently the most robust indication that neutrinos may
indeed have masses. With $m_{\nu_\mu}^2-m_{\nu_e}^2\approx 10^{-5}\,\rm eV^2$
(MSW solution, e.g.\ Hata and Haxton 1995) or $10^{-10}\,\rm eV^2$ (vacuum
solution, e.g.\ Krastev and Petcov 1994) all solar neutrino experiments can be
reconciled with each other and with theoretical flux predictions. This leaves
the possibility open that $\nu_\tau$ has a cosmologically interesting mass
which may show up, for example, in the $\nu_\mu$-$\nu_\tau$ oscillation
experiments CHORUS and NOMAD which are currently taking data at CERN (Winter
1995). Certain versions of ``see-saw'' neutrino mass models indeed suggest
that $\nu_\tau$ should carry a cosmologically significant mass if the MSW
solution of the solar neutrino problem obtains.

Of course, other more complicated neutrino mass schemes are conceivable, for
example one where the solar neutrino problem is solved by $\nu_e$-$\nu_\tau$
oscillations (e.g.\ Raffelt and Silk 1995). Such speculations are nurtured by
the candidate events at the LSND experiment which may be interpreted in terms
of $\anti\nu_\mu$-$\anti\nu_e$ oscillations with
$\Delta m^2_\nu\gapprox0.5\,\rm eV^2$, and perhaps a favored value of
$6\,\rm eV^2$ (Athanassopoulos et~al.\ 1995). While this interpretation is
controversial at the present time (Hill 1995), a confirmation with more
data would essentially establish that neutrinos do play a cosmologically
significant role.

\subsection{Neutralinos}

While neutrinos because of their small mass would constitute ``hot DM,'' a
plausible ``cold DM'' candidate is provided by supersymmetric extensions of
the particle-physics standard model.  Supersymmetry establishes a symmetry
between bosons and fermions so that in nature there should be a fermionic
partner to every bosonic degree of freedom and vice versa. Supersymmetry is
not realized among the known particles so that one predicts the existence of a
novel partner for every ``normal particle.'' They and supersymmetric ones
differ by a quantum number known as R-parity. If it is conserved, the lightest
supersymmetric particle (LSP) must be stable as required for a DM candidate.
Supersymmetry is thought to be a necessary ingredient of grand unification
theories (GUTs) which seek to unify the electromagnetic, weak, and strong
interactions at an energy scale of $f_{\rm GUT}\approx10^{16}\,\rm GeV$. The
main purpose of supersymmetry in GUTs is to explain the stability of the weak
interaction scale ($f_{\rm weak}\approx 250\,\rm GeV$) relative to
$f_{\rm GUT}$, and both relative to the Planck scale ($\mn1.2(19)\,\rm GeV$).
This is possible only if the masses of the supersymmetric particles are below
about $f_{\rm weak}$ so that their discovery is within the reach of current
and near-future particle accelerators.

A favored LSP that would double as a suitable DM candidate is the neutralino,
a linear combination of the supersymmetric partners of the photon, the $Z^0$
gauge boson, and the Higgs particle, which go by the name of photino, zino,
and higgsino, respectively. The neutralino's phenomenological properties are
closely related to those of neutrinos. One may think of a neutralino as a
Majorana neutrino (it is its own antiparticle) which interacts slightly weaker
than weak. The standard freeze-out calculation in the early universe then
yields a significant cosmic abundance if the mass is in the $3{-}300\,\rm GeV$
range.

The question if neutralinos are the DM of our galaxy can be addressed by
direct and indirect detection experiments. In the former one attempts to
measure the recoils of target nuclei hit by a galactic DM neutralino. The
expected event rate is very small (below $0.1$ per day and kg detector
material). While many efforts to search for this effect are pursued worldwide,
two experiments which are currently under construction deserve particular
mention. One (Cryogenic Rare Event Search with Superconducting
Thermometers---CRESST)
is built in the deep underground Gran Sasso laboratory (shielding from
cosmic rays!) by a collaboration between the
Max-Planck-Institut f\"ur Physik, M\"unchen, the Technische Universit\"at
M\"unchen, and Oxford University. The other (Cryogenic Dark Matter
Search---CDMS)
is built at a shallow site at Stanford by a collaboration organized
around the Center for Particle Astrophysics in Berkeley, California.  While in
the upcoming round of experiments the numerous free parameters of the
supersymmetric models must take on favorable values for neutralinos to
actually be detectable, it is noteworthy that these experiments for the first
time have a plausible chance of finding supersymmetric DM.

Neutralinos can be trapped in the Sun and Earth and annihilate there. The
annihilation products involve high-energy neutrinos which may be measurable in
the upcoming Cherenkov detectors AMANDA, DUMAND, NES\-TOR, and
Superkamiokande. Depending on details of the assumed supersymmetric model,
this ``indirect method'' may beat the direct search experiments to the
discovery of neutralinos.

For a detailed review and references concerning supersymmetric DM and methods
for its detection see Jungman, Kamionkowski, and Griest (1995).

\subsection{Axions}

Within quantum chromodynamics (QCD), the standard theory of strong
interactions, the neutron is expected to carry an electric dipole moment of
roughly the same magnitude as its magnetic one, while it is measured to be at
least a factor of $10^{-9}$ smaller. Because an electric dipole moment for any
fermion would violate the symmetry between particles and antiparticles
(CP~symmetry), its unexpected conservation by QCD is known as the ``CP~problem
of strong interactions.'' The most elegant solution invokes a new chiral U(1)
symmetry (Peccei-Quinn symmetry) which is spontaneously broken at some large
energy scale $f_a$. The corresponding Nambu-Goldstone boson is the axion, a
pseudoscalar particle which is closely related to the neutral pion. The axion
mass and interaction strength are roughly those of $\pi^0$, times $f_\pi/f_a$
where $f_\pi\approx 93\,\rm MeV$ is the pion decay constant. Because $f_a$ can
be very large, axions can be very light and very weakly interacting
(``invisible axions'').

In spite of their weak interactions axions are a QCD phenomenon. In the early
universe, they would be produced as coherent field oscillations during the QCD
phase transitions at $T\approx\Lambda_{\rm QCD}\approx150\,\rm MeV$. They are
produced in low-momentum modes, i.e.\ they are nonrelativistic from the start.
Thus, in spite of their small mass they are a cold DM candidate. In typical
scenarios the cosmic axion density is found to close the universe for
$m_a=10^{-5}{-}10^{-3}\,\rm eV$, roughly corresponding to
$f_a=10^{10}{-}10^{12}\,\rm GeV$. In contrast with neutrinos where
$\Omega_\nu h^2\propto m_\nu$ one finds for axions $\Omega_a h^2 \propto
f_a^{1.175}\propto m_a^{-1.175}$ so that axions with masses {\it below\/} the
closure limit are cosmologically excluded. On the other hand, axions with
relatively large masses (relatively small $f_a$ values) can be excluded
because they would be emitted from stars too efficiently to be consistent with
a variety of stellar evolution limits. Therefore, only a narrow ``window of
opportunity'' of roughly $10^{-6}\,{\rm eV}\lapprox m_a \lapprox
10^{-2}\,{\rm eV}$ remains where axions could still exist (Raffelt 1990,
1995). Put another way, if axions exist at all, and if the standard picture of
their coherent production in the early universe is correct, they must
contribute a significant fraction of the cosmic DM. This is a far more radical
statement than can be made about neutralinos because supersymmetry may well
exist in nature without providing the cosmic DM.

No method has yet been proposed that would allow one to discover ``invisible
axions'' in a pure laboratory experiment. However, galactic DM axions
can be searched by the ``haloscope'' method. The axion's
two-photon coupling allows for the radiative decay $a\to2\gamma$, and for the
``Primakoff conversion'' $a\leftrightarrow \gamma$ in the presence of an
external electric or magnetic field which plays the role of the second photon.
The conversion of galactic axions with, say, $m_a=10^{-5}\,\rm eV$ produces
microwaves of the same energy. Therefore, if a microwave resonator is placed
in a strong magnetic field one can search for the appearance of a narrow
line on top of its thermal noise signal.
One such experiment has recently taken up operation
in Livermore, California (van~Bibber et~al.\
1995) and another one is under construction
in Kyoto, Japan (Matsuki et~al.\ 1995). If axions are the DM of the
galaxy, this round of experiments has a first
realistic chance of finding them, in
contrast with two previous pilot experiments.

\subsection{Summary}

While it is easy to postulate some ad-hoc elementary particle that can serve
as a DM candidate, massive neutrinos, supersymmetric particles, and axions
have been invoked for other reasons and are thus well motivated. Current and
near-future search experiments may well discover these particles either in
pure laboratory experiments or in search experiments for galactic DM. It is
clear that one cannot truly establish nonbaryonic DM by indirect arguments, no
matter how compelling, unless one of the candidate particles with the
requisite properties is experimentally discovered. Therefore, the cosmological
importance of the DM search experiments, the accelerator searches for
supersymmetric particles, and the search for neutrino oscillations cannot be
overstated.

\section{Search for Dark Stars in our Galaxy}

Even though BBN indicates that baryons likely are not all of the DM, it also
indicates that likely there are more baryons in the universe than are visible
in the form of luminous matter. Therefore, independently of the final solution
of the DM problem, likely some baryons must hide themselves in the universe.
One obvious possibility is ionized intergalactic gas (Carr 1994). However,
baryons may also condense to form nonluminous stars and may contribute to the
DM in galactic halos. In principle, stellar remnants (white dwarfs, neutron
stars, black holes) and brown dwarfs are possibilities. A number of more or
less standard arguments have been advanced that brown dwarfs and perhaps
molecular clouds or black holes are the only plausible objects to hide baryons
in galactic halos and/or galaxy clusters (Carr 1994).

Small dark stars can be hunted in the halo of our galaxy by the gravitational
microlensing technique which relies on measuring the lightcurves of millions
of stars in the Large Magellanic Cloud over many years. Two collaborations
have observed first candidate events (Alcock et~al.\ 1995; Aubourg et~al.\
1995). At the present time it is not assured that these events can indeed be
attributed to lensing by galactic halo objects. If they were interpreted as
such, the small total number of events allows for any interpretation, i.e.\
that all or practically none of the halo mass is contributed by Massive
Astrophysical Compact Halo Objects (MACHOs). Thus, for now the most important
message is that the method works, and that with the collection of more data
the MACHO fraction of the halo mass will eventually emerge.

A purely baryonic halo for our galaxy may consist of brown dwarfs which form
something like dark globular clusters, and of molecular hydrogen clouds which
are yet another form to hide baryons. For various aspects of such a scenario
see De~Paolis, Ingrosso, Jetzer, and Roncadelli (1995). Better statistics from
future microlensing observations will reveal or exclude such a possibility.

\section{Black Holes}

Black holes are a DM candidate {\it sui generis} because they may form from
baryons, or they may be present in the universe as ``primordial black holes.''
For the purposes of structure formation, black holes which formed early enough
belong to the category of ``cold DM.'' For stellar-remnant black holes, severe
constraints exist which makes them appear implausible as DM candidates
(Carr 1994). Primordial black holes seem to be viable candidates, except that
the necessary abundance has to be taken on faith.

\section{Structure Formation}

The impressive isotropy of the CMB reveals a universe that was very
homogeneous at early times while the distribution of matter today is very
clumpy and structured. It is thought that the universe evolved from there to
here simply by the action of gravity, beginning with an initial spectrum of
low-amplitude density fluctuations that had been imprinted at some early
epoch, possibly during an inflationary phase. It is well known that this
scenario, if true, entails very restrictive bounds on the nature of~DM.

It is now a textbook wisdom that a purely baryonic universe is incompatible
with this scenario if the initial fluctuations were adiabatic so that
fluctuations in the baryon density imprint themselves directly on the CMB. If
the initial spectrum was of the isocurvature type the case against baryons is
strong but not as clear cut. In order to save such primordial isocurvature
baryon (PIB) models one must make use of the freedom of choosing an initial
power-law index for the fluctuation spectrum, and of choosing a suitable
ionization history of the universe which modifies the CMB characteristics.
Even with that much freedom such models are excluded unless one invokes a
cosmological term (Hu, Bunn, and Sugiyama 1995). Of course, a cosmological
term is now favored by many. However, in essence it constitutes a form of
nonbaryonic DM on a global scale, thus removing the main appeal of the PIB
scenario which involves only one form of DM without the need to explain
fine-tuned relative abundances of several components.

Even simpler structure-formation arguments exclude neutrino DM.  Because
neutrinos with $m_\nu\lapprox 30\,\rm eV$ stay relativistic until about the
radiation decoupling epoch, they can stream freely over large distances (``hot
dark matter''---HDM). Thus they wash out any previously imprinted density
fluctuation spectrum on small scales. With any plausible initial fluctuation
spectrum HDM is excluded except, perhaps, in a scenario where structure is
formed by cosmic strings or other primordial topological defects.

Apart from structure formation, neutrinos also have problems with small-scale
DM. A textbook argument involving the phase space of galactic halos as well as
Liouville's theorem shows that $m_\nu\gapprox 30\,\rm eV$ is required for
neutrino DM in a spiral galaxy like our own. For dwarf galaxies, something
like $m_\nu\gapprox 100\,\rm eV$ is needed, in contradiction with the
cosmological mass limit.

Weakly interacting particles that became nonrelativistic early (``cold dark
matter''---CDM) fare much better with regard to structure formation.
Neutralinos, or more general ``weakly interacting massive particles'' (WIMPs),
as well as axions fall into this category. The simplest CDM models have the
opposite problem of HDM in that they cause too much clumping of matter on
small scales.  There are a variety of solutions to this problem. The initial
fluctuation spectrum may be slightly ``tilted,'' i.e.\ not exactly of the
scale-invariant Harrison-Zeldovich form. Or there may be a small admixture of
neutrinos---a mass of $\sum m_\nu\approx5\,\rm eV$ seems to be ideal,
especially if it distributed equally among two or three flavors (e.g.\
Pogosyan and Starobinsky 1995).  Perhaps the most straightforward solution is
a model where CDM does not close the universe; it may be open, or it may be
critical by virtue of a cosmological term.

In summary, all conjectured DM candidates have problems with some structure
formation arguments. For now it looks as if for CDM these problems are the
least severe, or most easily patched up. Evidently it is not known if a
patched-up CDM cosmology or some completely different physical scenario
represents our universe.

\section{Cosmological Constant}

Vacuum energy of quantum fields can play the role of a cosmological
``constant'' $\Lambda$ which is then really interpreted as a dynamical
variable. Thus it is possible that $\Lambda$, while it is small or vanishing
today, was large in the very early universe and thus drove a de~Sitter
expansion (inflation). In this scenario $\Lambda$ must have evolved
dynamically to its present-day value. Its smallness remains unexplained,
and no compelling reason is known why $\Lambda$ today should be exactly zero
(Weinberg~1989).

Inflationary models of the universe have many virtues (B\"orner 1992; Peebles
1993). However, barring fine tuning they predict a vanishing spatial curvature
today which is not compatible with the measured values of other cosmological
parameters unless something like 65\% of the present-day cosmic energy density
resides in the $\Lambda$ term (Ostriker and Steinhardt 1995). On a global
scale DM may well consist of vacuum energy!

This hypothesis cannot be tested in the laboratory, but only by a careful
assessment of the parameters that characterize the
Robertson-Walker-Friedmann-Lema\^\i tre
models of the universe. Of particular importance is a
determination of the deceleration parameter---see Carroll, Press, and Turner
(1992) for a review, and Ostriker and Steinhardt (1995) for more recent
references. A serious theoretical prediction for the present-day value of
$\Lambda$, and notably if it has to be zero after all, likely must await the
emergence of a true quantum theory of~gravity.

\section{Alternatives to Dark Matter}

The hypothesis of particle DM requires nontrivial extensions of the
particle-physics
standard model. Thus it may seem no more radical to modify general
relativity (GR) such that there is no need for DM. In one phenomenological
approach (Modified Newtonian Dynamics---MOND; for a review
see Milgrom 1994) gravitational accelerations $a$ below a certain limit $a_0$
are given by $a^2/a_0=\GN M/r^2$. With $a_0\approx\mn1(-8)\,\rm cm\,s^{-2}$
this approach is surprisingly successful at explaining a broad range of DM
phenomena related to dwarf galaxies, spiral galaxies, and galaxy clusters
(Milgrom 1994, 1995).  Unfortunately, MOND lacks a relativistic formulation so
that it cannot be applied on cosmological scales.

One covariant alternative to GR is a conformally invariant fourth-order theory
(Mannheim 1995). In the nonrelativistic regime it leads to a linear
gravitational potential in addition to the Newtonian $1/r$ term. It explains
at least some of the galactic and cluster DM problems.

Before modifications of GR can be taken seriously they must pass relativistic
tests. An important case are galaxy clusters where large amounts of DM are
indicated by nonrelativistic methods (virial theorem) as well as by
relativistic indicators (gravitational lensing, notably giant arcs). Because
virial and lensing masses seem to agree well in several cases, scalar-tensor
extensions of GR are in big trouble, if not ruled out entirely (Bekenstein and
Sanders 1994).

Apparently, no serious attempt has been made to discuss truly cosmological
phenomena such as structure formation and CMB distortions in the framework of
alternate theories of gravity. At the present time it is not known if a
covariant theory of gravity exists that can explain the DM problems on all
scales.

\section{Summary}

Many solutions of the DM problem are on the table, but none is completely
convincing. A purely baryonic universe has the advantage that one does
not need to invoke hypothetical other components which miraculously have the
same cosmic abundance as baryons within a factor of order unity. It is far
easier to believe in some mechanism that hides, say, 90\% of all baryons in
some invisible form, say brown dwarfs, black holes, molecular gas, or an
ionized intergalactic medium. In detail, however, a baryonic scenario requires
extreme cosmological parameters, large systematic errors in BBN and/or the
observationally inferred primordial light element abundances, unmotivated
primordial isocurvature density fluctuations, and even then probably a
cosmological constant. One way out may be to hide enough baryons in black
holes which would effectively constitute CDM. However, the small-scale density
fluctuations that might cause the formation of black holes are not
well~motivated.

A slightly patched-up CDM cosmology fares quite well with regard to structure
formation. It is not known, however, if the correct particle-physics candidate
is among the ones which are currently favored (neutralinos and axions).
Extensions of the particle-physics standard model allow for the existence of
weakly interaction particles with the necessary properties, but do not by
themselves demand these properties. Particle CDM can be established only by a
direct or indirect measurement of the relevant candidate either in a pure
laboratory experiment, or as a constituent of the galactic halo. Several
current and near-future experimental efforts are beginning to address this
question in earnest.

If a $\nu_\tau$ mass of, say, $10\,\rm eV$ were discovered in ongoing
oscillation experiments, even HDM would have to be taken seriously again,
perhaps pointing to topological defects as a cause for structure formation. If
more data confirm the LSND claim of $\anti\nu_\mu$-$\anti\nu_e$ oscillations,
one would be led to believe that a HDM component is what patches up the CDM
cosmology.

Modified theories of gravity are able to address some DM problems, but so far
no covariant alternative to GR exists that would address all DM problems. Of
course, if a convincing such theory would emerge, this sort of explanation of
the DM problem would have to be re-assessed.

While arguments involving BBN and structure formation give us valuable and
quite compelling hints concerning the possible nature of DM, the question
likely cannot be settled without establishing a matter inventory of the Milky
Way by direct detection experiments which should turn up the right amounts of
particles, baryonic candidates, or black holes.


\section*{References}
\parindent=0pt
\parskip=0pt plus 1pt
\everypar={\hangindent=20pt\hangafter=1}
\frenchspacing

Alcock, C., et~al. (MACHO Collaboration). 1995. Experimental Limits on the
    Dark Matter Halo of the Galaxy from Gravitational Microlensing.
    {\sl Phys. Rev. Lett.} {\bf 74}:2867--2871.

Athanassopoulos, C., et~al. 1995. Candidate Events in a Search for
    $\anti\nu_\mu\to\anti\nu_e$ Oscillations. {\sl Phys. Rev. Lett.}
    {\bf 75}:2650--2653.

Aubourg, E., et~al. (EROS Collaboration). 1995. Search for Very Low-Mass
    Objects in the Galactic Halo. {\sl Astron. Astrophys.} {\bf 301}:1--5.

Bekenstein, J.~D., and R.~H.~Sanders. 1994. Gravitational Lenses and
    Unconventional Gravity Theories. {\sl Astrophys. J.} {\bf 429}:480--490.

B\"orner, G. 1992. The Early Universe---Facts and Fiction (2nd ed.).
    Berlin: Springer.

Carroll, S.~M., W.~H.~Press, and E.~L.~Turner. 1992. The Cosmological
    Constant. {\sl Ann. Rev. Astron. Astrophys.} {\bf 30}:499--542.

Carr, B. 1994. Baryonic Dark Matter. {\sl Ann. Rev. Astron. Astrophys.}
    {\bf 32}:531--590.

Carswell, R.~F., et al. 1994. Is there Deuterium in the $z=3.32$ Complex in
    the Spectrum of 0014+813? {\sl Mon. Not. R. Astron. Soc.}
    {\bf 268}:L1--L4.

Copi, C.~J., D.~N.~Schramm, and M.~S.~Turner. 1995. Big-Bang Nucleosynthesis
    and the Baryon Density of the Universe. {\sl Science} {\bf 267}:192--199.

De Paolis, F., G.~Ingrosso, P.~Jetzer, and M.~Roncadelli. 1995. Is the
    Galactic Halo Baryonic? {\sl Comm. Astrophys.} {\bf 18}:87--94.

Hata, N., and W.~Haxton. 1995. Implications of the GALLEX Source Experiment
    for the Solar Neutrino Problem. {\sl Phys. Lett. B} {\bf 353}:422--431.

Hata, N., et~al. 1995. Big Bang Nucleosynthesis in Crisis?
    {\sl Report} HEP-PH/9505319, {\sl Phys. Rev. Lett.}, to be published.

Hill, J.~E. 1995. An Alternative Analysis of the LSND Neutrino Oscillation
    Search Data on $\anti\nu_\mu\to\anti\nu_e$. {\sl Phys. Rev. Lett.}
    {\bf 75}:2654--2657.

Hu, W., E.~F.~Bunn, and N.~Sugiyama. 1995. COBE Constraints on Baryon
    Isocurvature Models. {\sl Astrophys. J.} {\bf 447}:L59--L63.

Jungman, G., M.~Kamiokowski, and K.~Griest. 1995. Supersymmetric Dark Matter.
    {\sl Phys. Rep.}, to be published.

Krastev, P.~I., and S.~T.~Petcov. 1994. New Constraints on Neutrino
    Oscillations in Vacuum as a Possible Solution of the Solar Neutrino
    Problem.  {\sl Phys. Rev. Lett.} {\bf 72}:1960--1963.

Malaney, R.~A., and G.~J.~Mathews. 1993. Probing the Early Universe: A Review
    of Primordial Nucleosynthesis Beyond the Standard Big Bang.
    {\sl Phys. Rep.} {\bf 229}:145--219.

Mannheim, P.~D. 1995. Linear Potentials in Galaxies and Clusters of Galaxies.
    {\sl Report\/} ASTRO-PH/9504022.

Matsuki, S., et al. 1995. Contribution to be published in:
    Proceedings of the XVth Moriond Workshop
    {\sl Dark Matter in Cosmology, Clocks, and Tests of Fundamental Laws},
    Villars-sur-Ollon, Switzerland, January 21--28, 1995.

Milgrom, M. 1994. Dynamics with a Nonstandard Inertia-Acceleration Relation:
    An Alternative to Dark Matter in Galactic Systems. {\sl Ann.\ Phys.\
    (N.Y.)} {\bf 229}:384--415.

Milgrom, M. 1995. MOND and the Seven Dwarfs. {\sl Report}
    ASTRO-PH/\break 9503056.

Ostriker, J.~P., and P.~J.~Steinhardt. 1995. The Observational Case for a Low-
    Density Universe with a Non-Zero Cosmological Constant. {\sl Nature}
    {\bf 377}:600--602.

Particle Data Group. 1994. Review of Particle Properties.
    {\sl Phys. Rev. D} {\bf 50}:1173--1826.

Peebles, P.~J.~E. 1993. Principles of Physical Cosmology. Princeton: Princeton
    University Press.

Pogosyan, D., and A.~Starobinsky. 1995. Mixed Cold-Hot Dark Matter Models With
    Several Massive Neutrino Types. Report ASTRO-PH/9502019,
    {\sl Astrophys. J.}, to be published.

Raffelt, G. 1990. Astrophysical Methods to Constrain Axions and Other Novel
    Particle Phenomena. {\sl Phys. Rep.} {\bf 198}:1--113.

Raffelt, G. 1995. Axions in Astrophysics and Cosmology.
    Proceedings of the XVth Moriond Workshop {\sl Dark Matter in Cosmology,
    Clocks, and Tests of Fundamental Laws}, Villars-sur-Ollon, Switzerland,
    January 21--28, 1995. {\sl Report} HEP-PH/9502358.

Raffelt, G., and J.~Silk. 1995. Can a Mass Inversion Save Solar Neutrino
    Oscillations from the LSND Neutrino? {\sl Report} HEP-PH/9502306,
    {\sl Phys. Lett. B}, submitted.

Rugers, M., and C.~J.~Hogan. 1995. {\sl Astrophys. J.}, submitted.

Songaila, A., L.~L.~Cowie, C.~J.~Hogan, and M.~Rugers. 1994. Deuterium
    Abundance and Background Radiation Temperature in High-Redshift Primordial
    Clouds. {\sl Nature} {\bf 368}:599--604.

Trimble, V. 1987. Existence and Nature of Dark Matter in the Universe.
    {\sl Ann. Rev. Astron. Astrophys.} {\bf 25}:425-475.

Tytler, D., and X.~M.~Fan. 1994. Deuterium and Metals at $z=3.57$ towards
    QSO 1937$-$1009. {\sl Bull. Am. Astron. Soc.} {\bf 26}:1424.

van Bibber, K., et al. 1995. A Second Generation Cosmic Axion Experiment.
    To be published in:
    Proceedings of the XVth Moriond Workshop
    {\sl Dark Matter in Cosmology, Clocks, and Tests of Fundamental Laws},
    Villars-sur-Ollon, Switzerland, January 21--28, 1995.
    {\sl Report} ASTRO-PH/9508013.

Weinberg, S. 1989. The Cosmological Constant Problem. {\sl Rev. Mod. Phys.}
    {\bf 61}:1--23.

Winter, K. 1995. Neutrino Oscillation Experiments at CERN.
    {\sl Nucl. Phys. B (Proc. Suppl.)} {\bf 38}:211--219.

\end{document}